\documentclass[
 aip,
 jap
 amsmath,amssymb,
 reprint,%
]{revtex4-1}

\usepackage{graphicx}
\usepackage{dcolumn}
\usepackage{bm}

\usepackage[utf8]{inputenc}
\usepackage[T1]{fontenc}
\usepackage{mathptmx}
\usepackage{etoolbox}
\usepackage{hyperref}
\usepackage{color}

\makeatletter
\def\@email#1#2{%
 \endgroup
 \patchcmd{\titleblock@produce}
  {\frontmatter@RRAPformat}
  {\frontmatter@RRAPformat{\produce@RRAP{*#1\href{mailto:#2}{#2}}}\frontmatter@RRAPformat}
  {}{}
}%
\makeatother
\begin{document}

\preprint{AIP/123-QED}
\title[Photoluminescence of Femtosecond Laser-irradiated Silicon Carbide]{Photoluminescence of Femtosecond Laser-irradiated Silicon Carbide}
\author{Y. Abdedou}
\affiliation{ 
RPTU University Kaiserslautern-Landau, Department of Physics and State Research
Center OPTIMAS, Kaiserslautern, Germany
 Erwin-Schroedinger-Strasse, 67663 Kaiserslautern. Germany 
}%
\author{A. Fuchs}
\author{P. Fuchs}
\affiliation{ 
Universität des Saarlandes, Fachrichtung Physik, Campus E2.6, 66123 Saarbrücken, Germany
}
\author{J. Heiler}
\affiliation{ 
Quantum Materials, Luxembourg Institute of Science and Technology (LIST), 28 Avenue des Hauts Fourneaux, 4362 Belval, Luxembourg
}
\affiliation{ University of Luxembourg, 2 Avenue de l'Université, 4365 Belval, Luxembourg
}
\author{D. Herrmann}
\affiliation{ 
Universität des Saarlandes, Fachrichtung Physik, Campus E2.6, 66123 Saarbrücken, Germany
}%

\author{S. Weber}
\author{M. Schäfer}
\affiliation{Institut für Oberflächen- und Schichttechnik GmbH, Research Center OPTIMAS, RPTU Kaiserslautern-Landau, 67663 Kaiserslautern Germany}%
\author{J. L'huillier}
\affiliation{Institut für Oberflächen- und Schichttechnik GmbH, Research Center OPTIMAS, RPTU Kaiserslautern-Landau, 67663 Kaiserslautern Germany}%
\author{F. Kaiser}
\affiliation{ 
Quantum Materials, Luxembourg Institute of Science and Technology (LIST), 28 Avenue des Hauts Fourneaux, 4362 Belval, Luxembourg
}
\affiliation{ University of Luxembourg, 2 Avenue de l'Université, 4365 Belval, Luxembourg
}
\author{C. Becher}
\affiliation{ 
Universität des Saarlandes, Fachrichtung Physik, Campus E2.6, 66123 Saarbrücken, Germany
}
\author{E. Neu}%
 \email{nruffing@rptu.de}
\affiliation{ 
RPTU University Kaiserslautern-Landau, Department of Physics and State Research
Center OPTIMAS, Kaiserslautern, Germany Erwin-Schroedinger-Strasse, 67663 Kaiserslautern. Germany 
}%

\date{\today}
\begin{abstract}
Silicon carbide (SiC) is the leading wide-bandgap semiconductor material, providing mature doping and device fabrication. Additionally, SiC hosts a multitude of optically active point defects (color centers) and is relevant for many applications in quantum technologies. 

A crucial step towards harnessing the full potential of the SiC platform includes technologies to create color centers with defined localization and density, e.g.\ to facilitate their coupling to nano-photonic structures and to observe cooperative effects. Here, silicon vacancy centers and divacancies stand out as no impurity atom is needed and high-thermal budget annealing steps can be avoided. We characterize the effect of localized, femtosecond laser irradiation of SiC, investigating surface modifications and photoluminescence including Raman spectroscopy and optical lifetime measurements. We employ commercial high-purity, semi-insulating substrates and an industrial grade laser system to explore broader applicability of the method. As a novel approach, we apply femtosecond laser irradiation to SiC substrates with an epitaxial graphene layer and find that the threshold for photoluminescence due to laser treatment is lowered.     \end{abstract}

\maketitle

\section{\label{sec:Introduction}Introduction}
Silicon carbide (SiC) has seen tremendous improvements as a semiconductor platform in recent years, especially in terms of crystal quality and wafer sizes that can now reach up to 8 inch. Together with these improvements, it has also been established as a host material for single color centers and their applications in quantum technologies. Silicon vacancy centers (V$_{Si}$) have several favorable characteristics for applications in quantum technologies. In 4H-SiC, they exist in two different configurations \cite{sorman_silicon_2000} corresponding to a missing silicon atom at a cubic and a hexagonal lattice site. The defects show zero phonon lines at 861 nm for \textit{V1} (cubic) and 916 nm for \textit{V2} (hexagonal) \cite{nagy_high-fidelity_2019, carter_spin_2015}. Moreover, V$_{Si}$ do provide an optically readable ground-state spin with a zero field splitting of 4.5 MHz for \textit{V1} and 70 MHz for \textit{V2} \cite{nagy_high-fidelity_2019, janzen_silicon_2009} which is addressable with standard RF components. However, the contrast of this optically detected magnetic resonance (ODMR) is comparably low (a few $\%$). An additional challenge is the fact that V$_{Si}$ luminescence is hard to collect in (0001) SiC, because of the V$_{Si}$ dipole orientation in the SiC lattice. (0001) material is however the material with the highest commercial availability. Milestones include super-radiance \cite{lukin_optical_2022} from two color centers, near-transform limited emission even in photonic nanostructures \cite{lukin_two-emitter_2023} as well as magnetic sensors working under extremely harsh conditions like high pressure \cite{wang_magnetic_2023}.
SiC has also emerged as a promising platform for epitaxial graphene growth by sublimation of SiC enabling the straightforward fabrication of large area, high quality graphene directly on a semiconducting substrate \cite{berger_ultrathin_2004, ohta_morphology_2008, emtsev_towards_2009}. Due to its high carrier mobility \cite{novoselov_room-temperature_2007} graphene has many possible applications in various fields \cite{novoselov_roadmap_2012} such as high frequency transistors \cite{lin_100-ghz_2010}, photonics \cite{tamura_fast_2022, rufangura_towards_2020} and also with V$_{Si}$ as transparent electrodes for Stark shift tuning of quantum systems \cite{ruhl_stark_2020}.

Localized V$_{Si}$ have been created by various techniques: ion implantation (focused ion beams \cite{wang_scalable_2017} or implantation through a mask \cite{wang_-demand_2019}) or laser writing \cite{chen_laser_2019, castelletto_photoluminescence_2018, zhou_silicon_2023}. Ion-based techniques in principle allow for the highest precision in localization which has been shown to be mainly limited by the straggle of the ions which is in the range of tens of nanometers for keV ion implantation \cite{pezzagna_nanoscale_2010}.  However, they also create high levels of damage and potential additional defects and require costly equipment. Laser writing has been investigated as an alternative that relies on highly non-linear processes. Previous works suggest the involvement of 16 photons in the generation of a vacancy defect in SiC \cite{chen_laser_2019}. With such an extremely non-linear process, even an optical technology allows for precise  localization that has been estimated to be as good as 80 nm\cite{chen_laser_2019}. 

To address different applications in quantum technologies, not only controlling the spatial placement but also varying the density of color centers is crucial: whereas single photon sources naturally require using a single color center, higher density ensembles can be favorable for highly sensitive magnetic sensors based on optically-readable spins. In this case, higher defect density enables brighter fluorescence and thus reduces photon shot noise in the optical read-out. Moreover,  enhancing defect density in ensembles may allow to observe the transition from uncoupled to collective emission (super-radiance) \cite{gutsche_revealing_2022}.  

Laser writing can be used to create single V$_{Si}$ in 4H-SiC as well as ensembles \cite{chen_laser_2019}. However, the creation of dense ensembles requires higher pulse energies that are well above the ablation threshold, leading to increased damages to the crystal \cite{wang_sem_2018}. In the context of localized creation of color centers, ablation processes are a parasitic effect and will limit the creation of  dense color center ensembles.
We here perform laser processing using a commercial laser structuring system irradiating commercial HPSI SiC. We elucidate if laser processing is compatible with using epitaxial graphene layers that have been suggested as transparent electrode to electrically tune color centers. We here report on the photoluminescence (PL) from localized laser irradiated areas, we analyze surface morphology, Raman spectra as well as PL spectra and lifetimes.

\begin{figure*}
\includegraphics[scale=0.85]{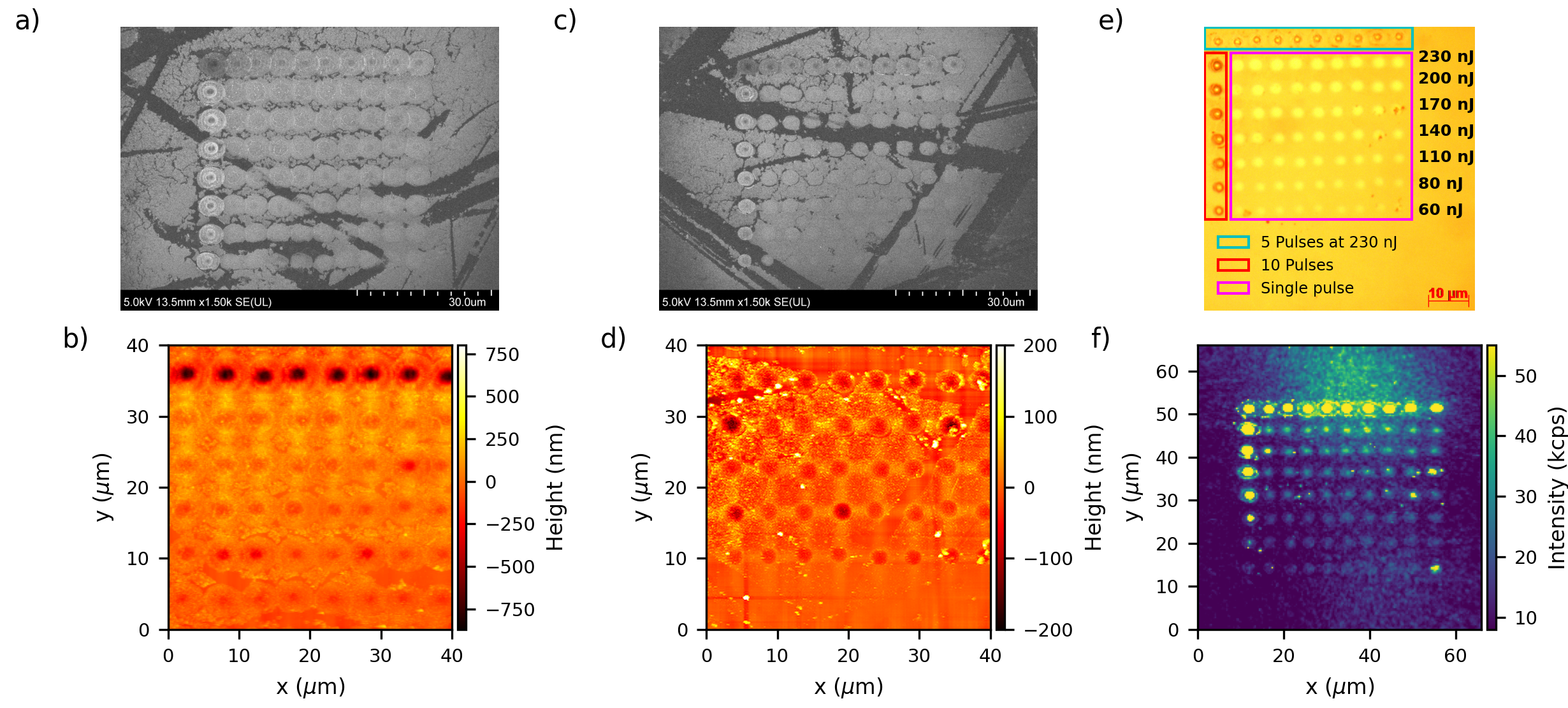}
\caption{\label{fig:graphene_SEM} \textbf{a)} and \textbf{c)} SEM images of laser written patterns 1 and 2 on the HPSI 4H-SiC sample with a graphene layer, with pulse energy ranging from 230-520 nJ (pattern 1) on \textbf{a)} and 60-230 nJ (pattern 2) on \textbf{c)}. \textbf{b)} and \textbf{d)} respective AFM measurements of the patterns in \textbf{a)} and \textbf{c)}.
\textbf{e)} White-light optical microscope image of the laser written pattern on the sample with graphene with higher pulse energy on the top row (230 nJ) and lower pulse energy on the bottom row (60 nJ) \textbf{f)} Associated confocal PL map of the laser written pattern on \textbf{c)} and \textbf{e)}.}
\end{figure*}
\section{\label{sec:sample_preparation}Sample pre-treatment}
 A 4-inch (0001) oriented HPSI 4H-SiC wafer purchased from Wolfspeed was diced into quadratic chips of < 10x10 mm$^2$. Diced chips were then solvent-cleaned with acetone and isopropanol to remove the protective resist layer applied for dicing. Surfaces are chemically mechanically polished to a roughness of < 0.5 nm according to manufacturer specifications. On the first sample, an epitaxial graphene layer was grown at 1600°C \cite{emtsev_towards_2009}. The second sample was annealed in vacuum (5.5 x $10^{-7}$ mbar) at 800°C for 3 hours. Annealing at temperatures above 600°C has been found to reduce the concentration of $V_{Si}$ \cite{widmann_coherent_2015, wang_-demand_2019} where ab initio calculations show that $V_{Si}$ migrate and recombine to form pairs of antisite vacancy ($C_{Si}V_C$) \cite{bockstedte_ab_2004, wang_formation_2013}.

\section{\label{sec:creation_graphene}Irradiation of HPSI with graphene layer}

We irradiate our 4H-SiC samples using an industrial grade laser structuring system (BlueCut, Menlo Systems GmbH, repetition rate 10 kHz)  with a wavelength of 1030 nm, a pulse duration of 383 fs and a repetition rate up to 1 MHz at a maximum pulse energy of 10 µJ. We focus the beam using a Thorlabs LMH-20X-1064 objective with 0.4 NA. The laser beam is linearly polarized. The focused beam had a calculated diameter of 2.4 $\mu$m ($1/e^2$) in air. The sample was then cleaned in acetone in an ultrasonic bath to remove particles collected on the surface during processing.  The laser-written patterns labeled pattern 1 and 2 were imaged using a commercial scanning electron microscope (SEM, Hitachi SU8000) in figure \ref{fig:graphene_SEM} \textbf{a)} and \textbf{c)}, and with a white-light optical microscope in figure \ref{fig:graphene_SEM} \textbf{e)}, corresponding to the pattern shown in figure \ref{fig:graphene_SEM} \textbf{c)}. The top row and the leftmost column serve as markers, on pattern 1 we used 5 pulses per spot at 520 nJ on figure \ref{fig:graphene_SEM} \textbf{a)} (230 nJ on figure \ref{fig:graphene_SEM} \textbf{c)} for pattern 2), and 10 pulses per spot with pulse energy ranging from 230 nJ to 520 nJ on figure \ref{fig:graphene_SEM} \textbf{a)} for pattern 1 (60 nJ to 230 nJ on figure \ref{fig:graphene_SEM} \textbf{c)} for pattern 2) for the leftmost column to induce a clear surface modification. The rows of spots visible within this marker angle were processed using a single pulse of the laser with pulse energies ranging from 230 nJ to 520 nJ on figure \ref{fig:graphene_SEM} \textbf{a)} for pattern 1 (60 nJ to 230 nJ on figure \ref{fig:graphene_SEM} \textbf{c)} for pattern 2). This preliminary investigation by SEM and white-light microscopy already indicates a challenge in laser writing of (dense) color center ensembles: In the regions where multiple pulses have been used for processing, around the laser processed area a structure occurs that is due to laser ablation and re-deposition of material. This re-deposited material however will not be of high quality and is typically observed to be amorphous \cite{zhou_silicon_2023}. 

\begin{figure}
\includegraphics[scale = 0.85]{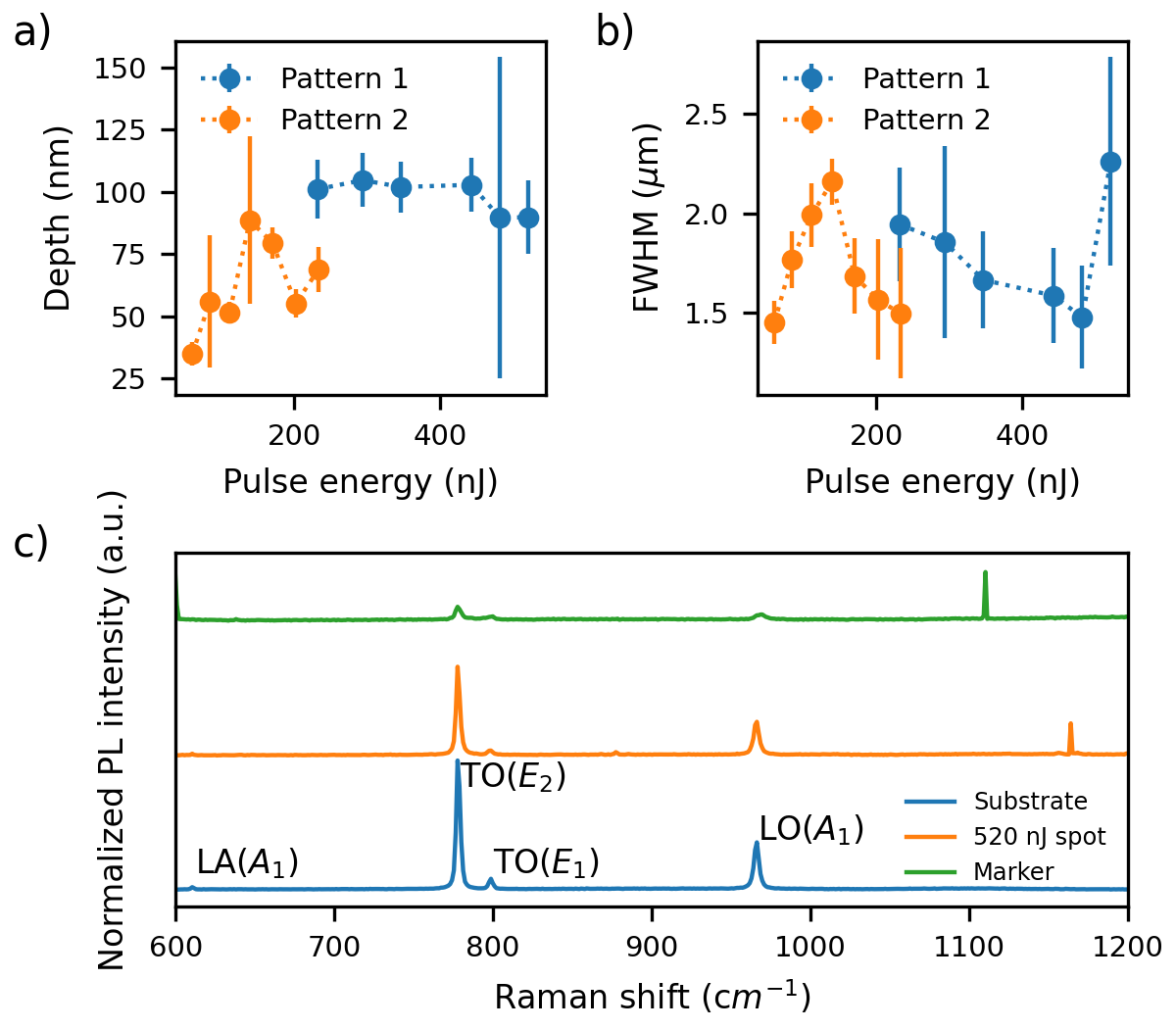}
\caption{\label{fig:graphene_depth_width} \textbf{a)} and \textbf{b)} AFM measured depth and width (FWHM) of the laser written spots on the HPSI 4H-SiC sample with a graphene layer.\textbf{c)} Raman spectra of the unprocessed HPSI 4H-SiC with graphene, a spot irradiated using a 520 nJ pulse and a marker. Spectra are offset vertically for clarity.}
\end{figure}

We assume that light gray areas in figure \ref{fig:graphene_SEM} \textbf{a)} and \textbf{c)} are graphene while darker parts are attributed to be the bare 4H-SiC surface. In darker areas, graphene on the surface appears to be removed, likely occurring during laser processing when ablated material was ejected scratching the surface. Additional damage may have occurred during sample handling. Although laser written spots exhibit a contrast similar to that of the surrounding graphene layer, this does not necessarily indicate that graphene remains at these sites, as all pulse energies used are well above the ablation threshold of graphene\cite{dong_evaluating_2016, roberts_response_2011}. Instead, their higher brightness compared to the bare substrate may arise from surface modification/ablation of 4H-SiC induced by laser pulses, as observed in femtosecond laser ablation of 4H-SiC studies \cite{chen_effect_2022, wang_experimental_2022}. To further study surface modifications on the graphene and 4H-SiC, atomic force microscope (AFM) measurements were performed on both patterns 1 and 2 using a commercial AFM (Park AFM XE-70). AFM images are shown in figures \ref{fig:graphene_SEM} \textbf{b)} and \textbf{d)} which are associated to laser written patterns 1 and 2 in \ref{fig:graphene_SEM} \textbf{a)} and \textbf{c)}. The femtosecond laser pulses created craters on the 4H-SiC surface as observed in literature \cite{castelletto_photoluminescence_2018}. On figure \ref{fig:graphene_depth_width} \textbf{a)} and \textbf{b)}, we measured the depth and width (FWHM) of the craters created during the laser processing. We observe different trends for the the higher pulse-energy pattern (pattern 1) and lower pulse-energy pattern (pattern 2). On the lower energy pattern, depth and width increase up to 140 nJ and decrease until 200 nJ for the depth and 230 nJ for the width. At 230 nJ, the average depth is 40 nm bigger in the higher pulse energy pattern and the spots are 0.5 $\mu$m wider. On the other hand, in the higher energy pattern, spot depth remains relatively constant with increasing pulse energy, while spots width tends to decrease with pulse energy until 480 nJ. We assume that these differing evolutions of spot dimensions as a function of pulse energy between the two patterns arise from variations in laser focusing while moving the sample to different writing locations.
Figure \ref{fig:graphene_depth_width} \textbf{c)} shows Raman spectra taken in a home-built confocal laser-scanning fluorescence microscope (CLSM), labeled CLSM 1. In CLSM 1, the sample was cooled down to 4.3 K in a closed-cycle cryostat (CryoVac). In CLSM 1, we excite the SiC sample using a tunable titanium sapphire laser (Sirah Matisse) operating at 737.19 nm focused on the sample with a microscope objective (Olympus MPLN 100x, NA = 0.9). PL and Raman scattered light is collected using the same objective and reflected laser light is rejected using a 740 nm dichroic mirror. Transmitted PL is further filtered with a longpass filter and coupled to a single mode fiber that can either be connected to an APD (Excelitas SPCM-AQRH) or a spectrometer (Horiba iHR550) equipped with a 600 lines per mm grating and a Peltier cooled CCD camera.
Measurements were performed on an unprocessed part of the sample, a written spot at 520 nJ and a marker made with a train of 10 pulses at 520 nJ. The Raman lines observed on the substrate are characteristic of 4H-SiC. At 520 nJ pulse energy, the written spots exhibit Raman lines with intensities comparable to the labeled peaks from the unprocessed substrate. Whereas, compared to the substrate, TO($E_2$), LO($A_1$) and TO($E_1$) peaks from the marker have smaller intensities and are broadened, TO($E_1$) is shifted and LA($A_1$) is not visible, indicating severe damages caused by the train of 10 pulses. While this Raman spectrum shows no sign of amorphous material even for 1850 nJ, the strong decrease in Raman peak intensity has been associated with significant crystal damage in previous work \cite{castelletto_photoluminescence_2018}. The Raman spectrum does not show clear evidence for remaining graphene in accordance with the observation of ablation of material at the surface.  
   
To investigate photoluminescence from the laser processed area, we use another home-built CLSM labeled CLSM 2. In CSLM 2, we perform room temperature characterization. We use continuous 785 nm laser light (Coherent OBIS LX 785) for excitation. In addition, we can couple a tunable, pulsed super-continuum laser source (NKT Super K-Extreme, typical pulse width < 100 ps) which we use to measure the excited state lifetime $\tau$ of the observed PL. In both cases, we use a 100x microscope objective with 0.9 NA (Zeiss EC EPN 100x/0,9 DIC) to focus the laser onto the sample and to collect PL in a reflection geometry. We separate PL and reflected laser light using a 785 nm edge dichroic mirror (Semrock Laser-Beamsplitter HC R785 lambda/5 PV flat). We detect PL using silicon avalanche photo-diodes (APDs) for spectrally-integrated detection (Excelitas SPCM-AQRH-14). Alternatively, we sent the PL to a spectrometer equipped with a silicon CCD (Princeton Instruments Acton Standard SP-2558 Spectrometer with Pixis 256E CCD). In both cases, we use a 850 nm long pass filter for additional laser filtering. Figure \ref{fig:graphene_SEM} f) shows a room temperature confocal PL map of pattern 2 from figure \ref{fig:graphene_SEM}. c). Localized PL is visible resembling the regular array that was processed by irradiating the SiC sample with single laser pulses, however, a significant level of PL also originates from areas not irradiated by the laser due to already present native defects remaining in our substrate, the PL ratio between the written spots at 60 nJ (bottom-most row) with the background is only \textasciitilde 2.

\section{\label{sec:creation}Irradiation of pristine HPSI SiC}
\begin{figure*}
\includegraphics{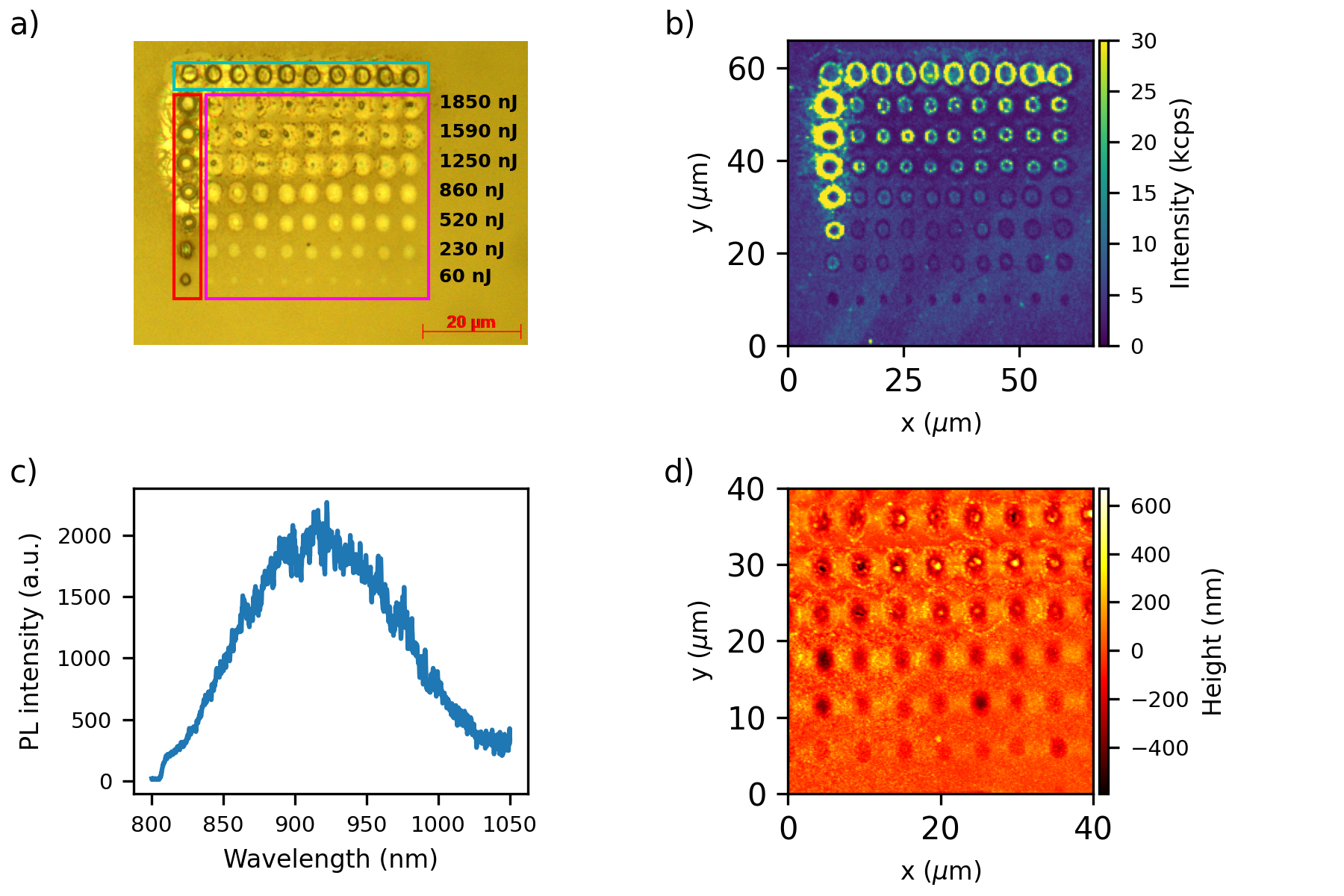}
\caption{\label{fig:pristine_confocal_spectra} \textbf{a)} White-light optical microscope image of the laser-processed pattern observed on the surface. The top row and the leftmost column serve as a marker, here we used a train of 5 pulses per spot at 1850 nJ for the top row (cyan box), and 10 pulses per spot with pulse energy ranging from 60 to 1850 nJ for the leftmost column to induce a clear surface modification (red box). The rows of spots visible within this marker angle have been processed using a single pulse of the laser with pulse energies ranging from 60 - 1850 nJ (pink box). \textbf{b)} Confocal PL map of the laser written pattern presented in FIG. \ref{fig:pristine_confocal_spectra} a). PL with wavelengths > 850 nm was detected. From top to bottom, the rows have been written with 1850, 1590, 1250, 860, 520, 230 and 60 nJ (single pulse per spot).\textbf{c)} Exemplary PL spectrum from a spot irradiated with 860 nJ. \textbf{d)} AFM image of the laser written pattern.}
\end{figure*}

As a reference, we treated a pristine HPSI 4H-SiC substrate using the same method described in the previous section, except using for this sample a repetition rate of 100 kHz to make the markers: we wrote a pattern on the surface with pulse energies ranging from 60 nJ (bottom row) to 1850 nJ (top row).  Figure \ref{fig:pristine_confocal_spectra} a) shows a white-light optical microscope image of the resulting pattern.
Figure \ref{fig:pristine_confocal_spectra} b) shows a confocal PL map of the laser processed area. The bottom-most row has been exposed to 60 nJ laser pulses, here, PL intensity is even slightly reduced compared to the surrounding, unprocessed SiC. This could be attributed to local annealing \cite{wang_femtosecond_2024} and surface modification as our 60 nJ pulses have energies above the reported modification threshold \cite{chen_effect_2022} but below the ablation threshold \cite{wang_experimental_2022} necessary to create vacancies. 
\begin{table*}[]
    \centering
    \begin{tabular}{|c|c|c|c|c|c|c|}
        \hline
        \textbf{Reference} &  \boldsymbol{$\lambda$} & \boldsymbol{$t_p$} & \textbf{Laser Focusing Objective} & \boldsymbol{$E_P$} & \boldsymbol{$I_P$} & \textbf{Comments}\\
        \hline
        \cite{chen_laser_2019} & 790 nm & 250 fs & 1.4 NA 60x oil & 10.7 nJ & 22.2 TW/cm$^2$ &Threshold for creating single $V_{Si}$\\
        \cite{castelletto_color_2020} & 1030 nm & 230 fs & 0.9 NA 100x & 58 nJ & 30.2 TW/cm$^2$ &Threshold for PL creation in the treated area\\
        This work & 1030 nm & 383 fs & 0.4 NA 20x & 60 nJ & 5 TW/cm$^2$ &Lowest pulse energy used, onset of PL for 4H-SiC with graphene\\
        This work & 1030 nm & 383 fs & 0.4 NA 20x & 230 nJ & 19.2 TW/cm$^2$ &Onset of PL for pristine 4H-SiC\\
        \hline
    \end{tabular}
    \caption{Summary of femtosecond lasers and optics parameter used in this works and for various previous work. $t_p$ is the pulse duration, $E_P$ the pulse energy and $I_P$ the corresponding peak intensity. }
    \label{tab:laserparameters}
\end{table*}
Ref.\ \cite{chen_laser_2019} identifies the process that creates $V_{Si}$ in SiC as highly non-linear. According to Ref.\ \cite{chen_laser_2019}, 16 photons at a wavelength of 790 nm are involved in creating a single $V_{Si}$ defect. For such a highly non-linear process (among other parameters), the peak intensity of the laser pulse will be crucial. We thus compare the laser and optics parameters in our work to previous work in table \ref{tab:laserparameters}. We find a slightly lower threshold than previously reported for the onset of PL on the pristine sample and a strongly reduced threshold for the sample with graphene. This could be explained by the absorptance of graphene which is 2.3 \% \cite{nair_fine_2008} compared to only 0.16\% at its surface in 4H-SiC at 1030 nm\cite{yan_crater-shaped_2024} making it 14 times higher.  

\begin{figure}
\includegraphics[scale = 0.85]{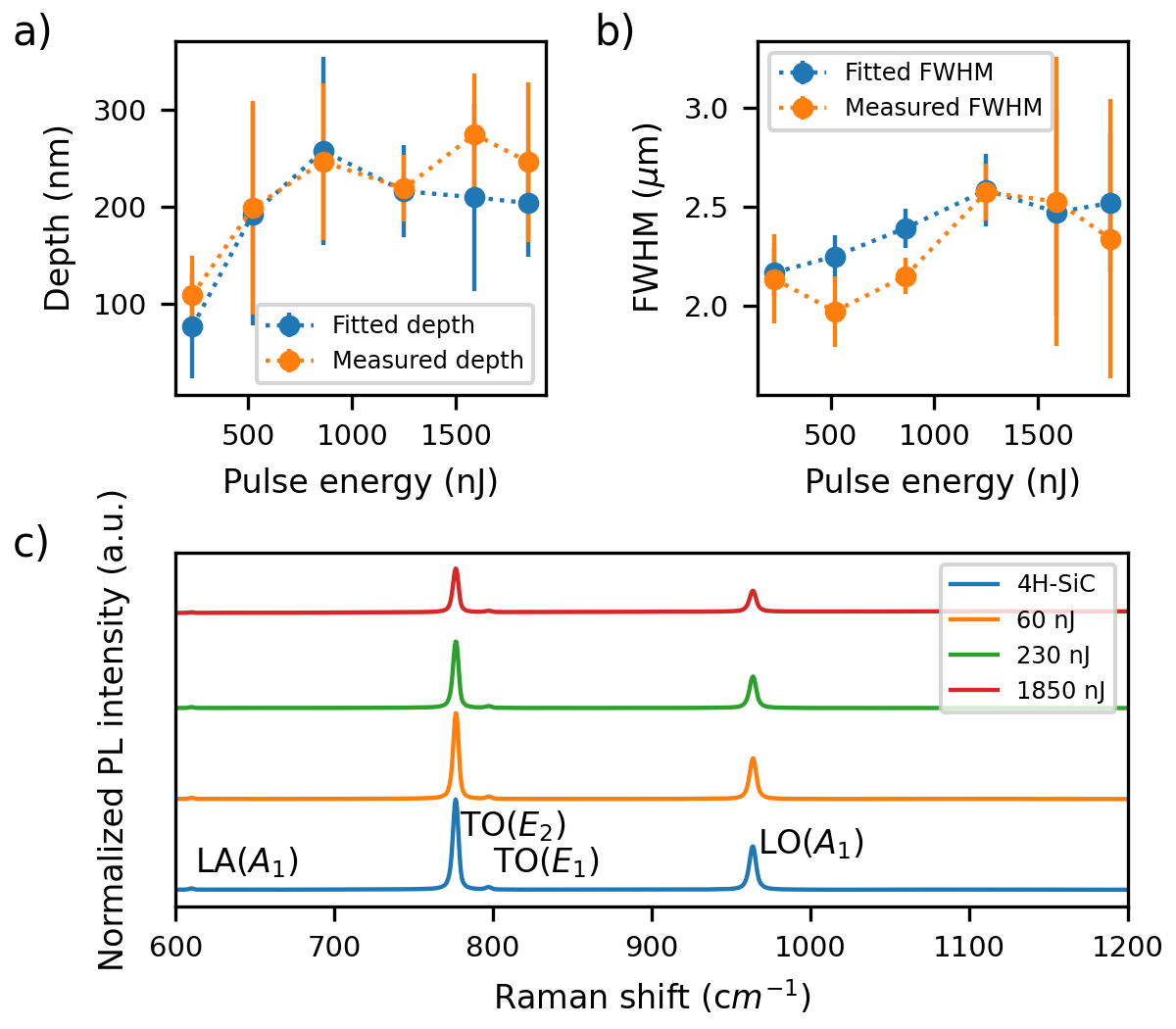}
\caption{\label{fig:pristine_raman_afm}   \textbf{a)} Average depth of the laser written spots in function of the laser pulse energy measured on spots profiles in orange and given by a fit with a Gaussian in blue. \textbf{b)} Average width (FWHM) of the laser written spots in function of the laser pulse energy. \textbf{c)} Raman spectra of the unmodified pristine HPSI 4H-SiC and of spots made at different laser pulse energies of 60 nJ, 230 nJ and 1850 nJ. The spectra are offset vertically for clarity and have been recorded using a 532 nm laser.}
\end{figure}
Figure \ref{fig:pristine_raman_afm}. c) shows Raman spectra taken using a commercial Raman spectrometer with a 532 nm excitation laser on the pristine sample from an unprocessed area and from spots written at 60 nJ, 230 nJ and 1850 nJ. Raman spectra taken from dimples made with single pulse at 60 nJ show reduced peak intensities most notably on the TO($E_2$) and LO($A_1$) peaks compared to unprocessed 4H-SiC. At 230 nJ and 1850 nJ, the decrease in peak intensities is even more significant, being maximal at 1850 nJ, again indicating significant damage to the crystal lattice. 
To further analyze the structures, we again employ AFM measurements. The resulting image in figure \ref{fig:pristine_confocal_spectra}. d) shows spots made with single pulses from 230 nJ to 1850 nJ. For the spots irradiated with 60 nJ, a shallow surface modification is visible. For spots written with pulse energy between 230 nJ to 860 nJ ablated material results in a crater-like hole. At higher pulse energy, we find a dimple with a hillock in the middle and ring shaped wall in the outer area. Figure \ref{fig:pristine_raman_afm} a) and b) summarize the depth and width of the surface structures. We add two datasets in each graph: we estimate the depth by measuring the distance between the surface and the deepest spot in the structure and we approximate the structure using a Gaussian fit. As discernible from figure \ref{fig:pristine_raman_afm} a) and b) this gives us similar absolute values and a similar trend: size and width of the surface structures roughly grow with pulse energies up to 750 nJ. The diameter (width at 1/$e^2$) of the features obtained above 1250 nJ with roughly 2.5 $\mu$m matches the diameter of the laser beam's Gaussian intensity profile that is 2.4 $\mu$m.  However, for pulse energies above 860 nJ, deviations from a Gaussian shape occur.

\begin{figure}
\includegraphics[scale = 0.85]{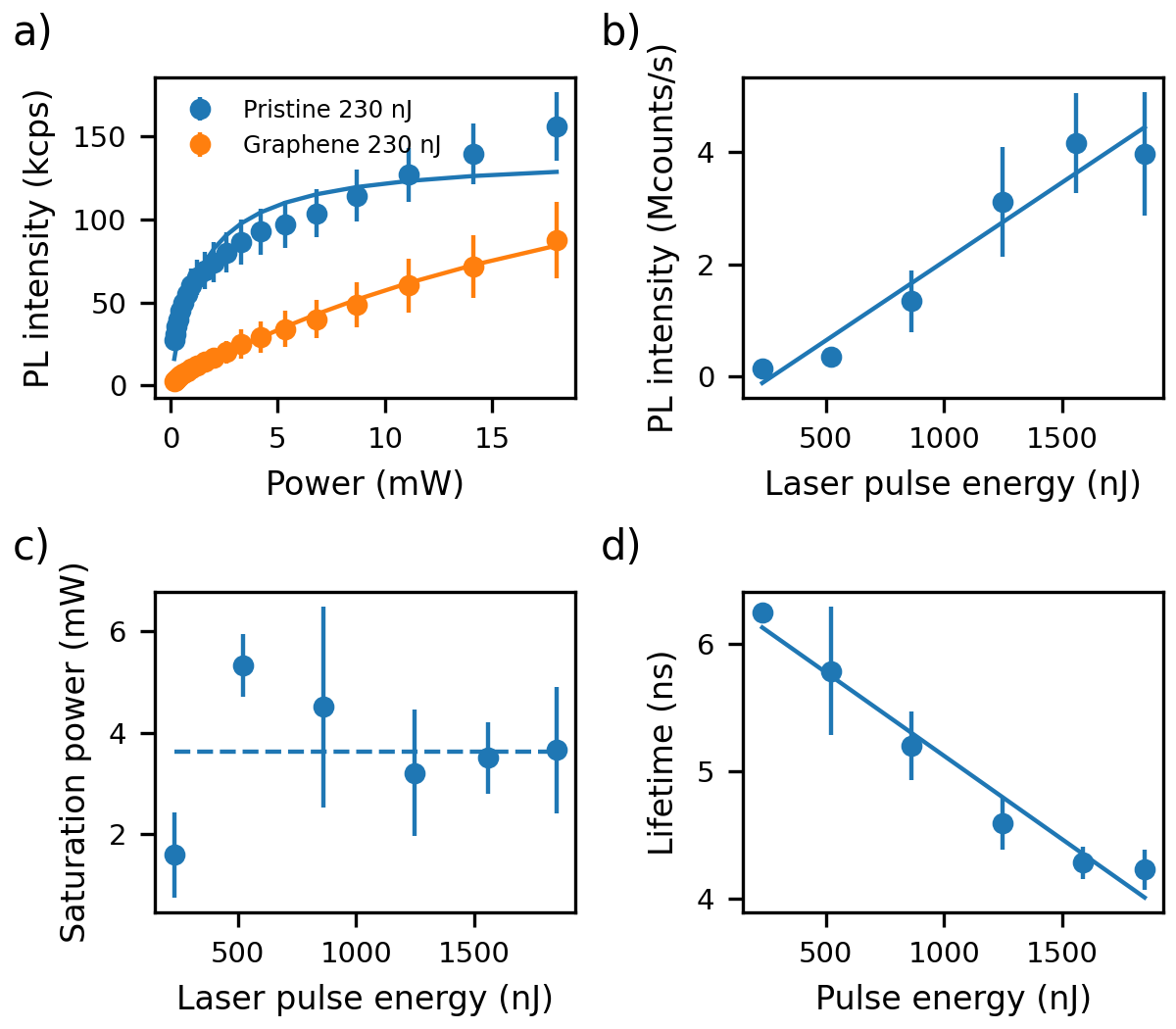}
\caption{\label{fig:saturation_lifetime} \textbf{a)}  Comparison of saturation of PL induced via laser processing on the pristine 4H-SiC and 4H-SiC with a graphene layer of spots written with a 230 nJ pulse energy.  \textbf{b)} Saturation count rate detecting PL with wavelengths longer than 850 nm in function of laser pulse energy. \textbf{c)} Saturation power associated with the saturation count rate in function of the pulse energy, dashed line shows the mean saturation power at 4.72 mW. \textbf{d)} Lifetime measurements detecting PL with wavelengths longer than 850 nm in function of the pulse energy. }
\end{figure}

We then perform PL spectroscopy of the bright spots created by irradiation with single laser pulses. For all photo-luminescent spots, we obtain typical room temperature spectra as displayed in figure \ref{fig:pristine_confocal_spectra} c). This very broad PL spectrum without a discernible zero-phonon line has been identified as the typical PL signature of $V_{Si}$ at room temperature \cite{carter_spin_2015}. The measured PL intensities and the fact that second order photon correlation measurements on the bright spots show no anti-bunching point defect ensembles. Room temperature spectra are similar to previous work \cite{castelletto_photoluminescence_2018}.


We measure PL saturation curves for each laser irradiated spot and average over the measurements on the spots irradiated with the same pulse energy. Figure \ref{fig:saturation_lifetime} a) exemplary shows the saturation measurement for spots irradiated with 230 nJ. To correct the data for background, we perform a saturation measurement at a spot 10 $\mu$m downward from the row written at 60 nJ. We correct the data points by subtracting the background. We note that the PL background also showed a saturation behavior. This might indicate that native point defects contribute to the background as reported previously \cite{castelletto_color_2020}. Figure \ref{fig:saturation_lifetime} b) shows the saturation count rate as a function of the writing pulse energy. Saturation count rates increase linearly with laser pulse energy indicating a linear increase of luminescent defects with pulse energy assuming a constant emission rate for each created defect. On the other hand, we will below discuss changes on excited state lifetimes that indicate non-radiative processes and thus changes in the emission rates. 

In addition, modifying the surface profile (Figure \ref{fig:pristine_raman_afm}), especially for increased pulse energy also changes the crystalline properties of SiC in the vicinity of the spots \cite{zhou_silicon_2023, liu_confocal_2020, castelletto_photoluminescence_2018}. When aiming to create $V_{Si}$,  carbon vacancies ($V_C$) are also created nearby due to their low formation energy. Carbon vacancies are not optically active. They provide non-radiative decay paths for $V_{Si}$ therefore quenching their PL \cite{steeds_transmission_2002, kraus_three-dimensional_2017, chen_laser_2019}. Figure\ref{fig:saturation_lifetime}. d) summarizes the lifetimes $\tau$ extracted from PL measurements (> 850 nm). Again, we obtain by averaging measurements from all spots irradiated with the same pulse energy. For spots irradiated with 230 nJ pulses, we find $\tau = 6.2$ ns which matches $\tau_{V_{Si}} = 6.1$ ns obtained for $V_{Si}$ created using electron irradiation \cite{hain_excitation_2014} again hinting at the creation of $V_{Si}$ . $\tau$ linearly decreases with pulse energy, indicating quenching by $V_C$ and also potential interaction with extended defects due to laser irradiation. In contrast, the saturation power as displayed in figure \ref{fig:saturation_lifetime} c) did not change with pulse energy.

\begin{figure}
\includegraphics[scale = 0.85]{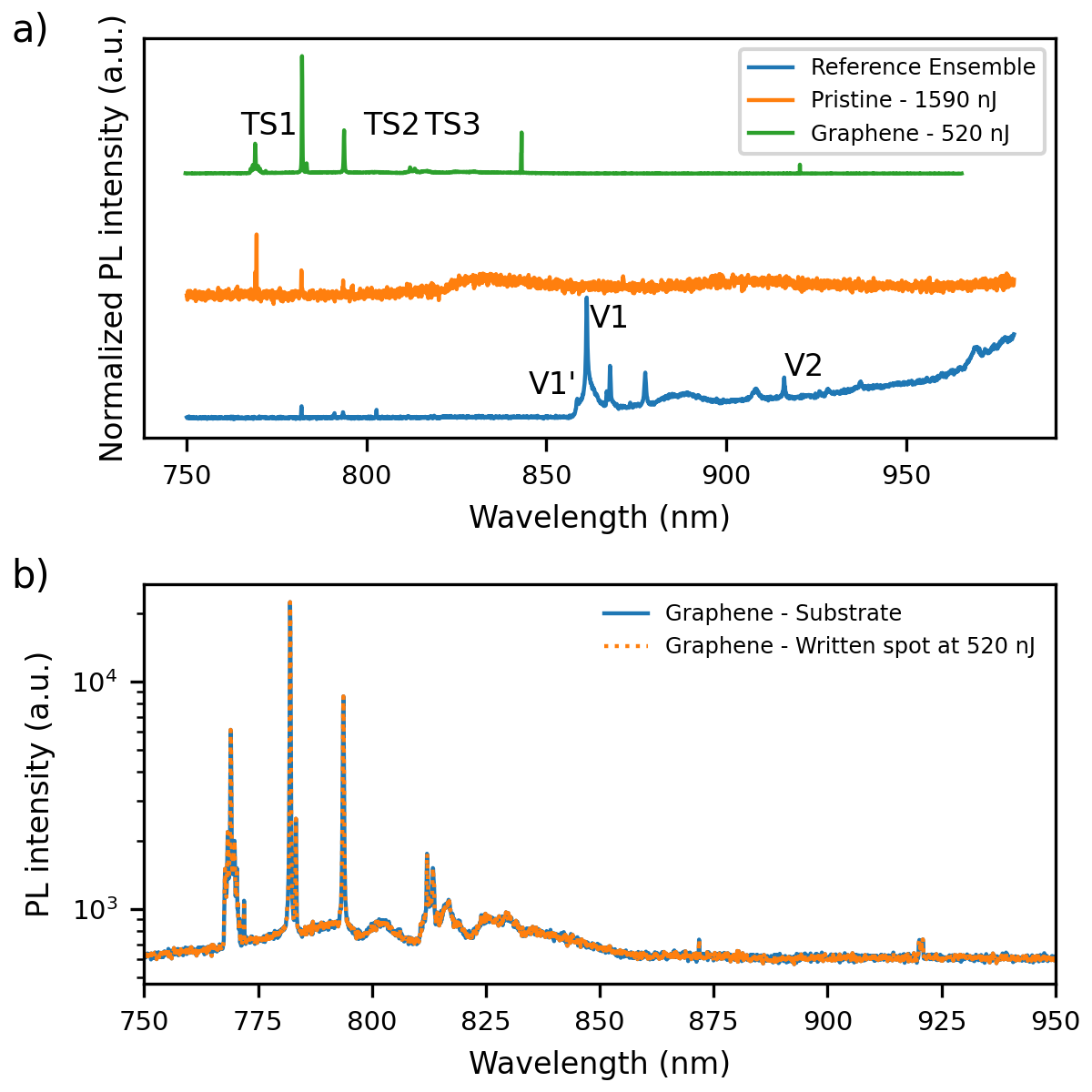}
\caption{\label{fig:lowtemp_comparision} \textbf{a)} PL spectra at 4.3K taken on a reference ion implanted sample which is compared to a laser written spot at 1590 nj on the pristine sample and a laser written spot at 520 nJ on the sample with graphene. Peaks associated with $V_{Si}$ are labeled V1' (858 nm), V1 (861 nm) and V2 (916 nm). Peaks associated to the TS center are labeled TS1 (769 nm), TS2 (812 m) and TS3 (813 nm). \textbf{b)} Low temperature
spectra (4.3K) of the sample with graphene taken on an unprocessed area (blue) and a laser written spot (520 nJ) (orange).}
\end{figure}
\begin{figure}
\includegraphics[scale = 0.85]{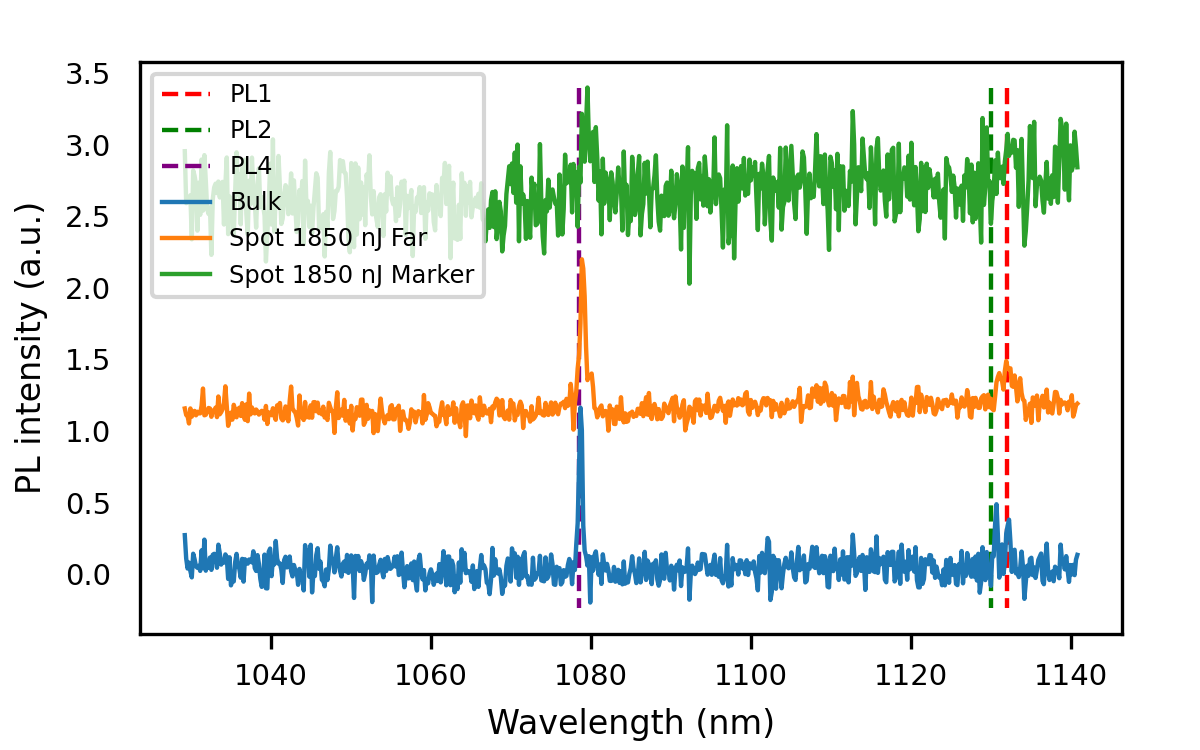}
\caption{\label{fig:lowtemp_NIR_Pristine} PL spectra at 5K taken on the pristine 4H-SiC sample at different position, unprocessed region (blue), around a written spot at 1850 nJ far from the markers (orange) around a written spot at 1850 nJ close to the markers (green).}
\end{figure}

To obtain more information on the defects created by the laser writing process, both samples (graphene and pristine) and a reference HPSI 4H-SiC sample implanted with 350~keV protons at a dose of $1 \times 10^{15}\,\mathrm{cm^{-2}}$ were cooled down to 4.3K in the CLSM 1 setup described in the previous section.

Figure \ref{fig:lowtemp_comparision} a) compares the cryogenic spectra of three 4H-SiC samples: ion-implanted reference sample, laser written pristine sample, and laser written sample with a graphene layer. The reference sample exhibits peaks associated with silicon vacancies ($V_{Si}$), including V1' at 858 nm, V1 at 861 nm, and V2 at 916 nm, in agreement with literature.

While our room-temperature spectra and lifetime measurements strongly suggested the localized formation of $V_{Si}$ (see figure \ref{fig:pristine_confocal_spectra} d) and figure \ref{fig:saturation_lifetime} d)), we do not observe ZPL emission from $V_{Si}$ at cryogenic temperatures in both laser-written samples. This suggests that efficient, localized creation of $V_{Si}$ is not feasible under our experimental conditions and using HPSI material. Our findings align with those reported in \cite{castelletto_color_2020}, where $V_{Si}$ centers were detected in laser-written regions but only with a weak enhancement relative to the non-irradiated substrate. Moreover, as summarized in table \ref{tab:laserparameters} previous laser writing work performed at the same 1030 nm wavelength did not show ZPL related to $V_{Si}$, only PL creation and typical room temperature spectra were observed. $V_{Si}$ centers occurring before laser writing in our substrates were likely eliminated due to the high-temperature annealing for the pristine sample and the graphene growth for the other sample.

The PL spectrum of the sample with graphene exhibits peaks corresponding to the TS center at 769 nm, 812 nm, and 813 nm (TS1, TS2, and TS3, respectively) \cite{ruhl_controlled_2018}. To determine whether TS centers form during laser writing or graphene growth, we analyzed spectra from a laser-written spot (520 nJ) and an unprocessed region of the substrate, as shown in figure \ref{fig:lowtemp_comparision} b). The absence of significant spectral differences suggests that TS centers originate from the graphene growth process and not from laser irradiation. Additionally, the Raman spectra of both regions exhibits comparable intensity and linewidth, indicating similar levels of crystal damage.
The TS defect is known to withstand high annealing temperatures, but its microscopic origin remains uncertain and TS centers cannot serve as indicators for studying the laser-writing process. The absence of $V_{Si}$ in laser-written regions may suggest that  high temperatures in the processed volume lead to their annealing. Moreover, for $V_{Si}$ centers to exhibit PL, they must be in their negatively charged state. Charge-state alterations caused by the presence of other defects may therefore suppress $V_{Si}$ PL emission.

Moreover, we investigated the near-infrared (NIR) PL using a dedicated setup optimized for detection above 1000 nm. Within this wavelength range falls the PL of divancancy defect in SiC. The samples were cooled down to 5 K in an attoDRY800 cryostat, and different excitation wavelengths were employed: 730 nm and 915 nm from a Sirah Ti:Sa Matisse CR laser, and 810 nm from a Toptica diode (EYP-RWL-0808-00800-4000 BFW09-0000). The excitation beams were focused onto the samples and PL was collected using a Zeiss 100×, NA 0.9 objective. Reflected laser light was removed from the collected PL using dichroic mirrors and a set of long-pass filters before being sent to a SpectraPro HRS 750 spectrometer equipped with a NIRvana HS camera.

The sample with a graphene layer was investigated under 810 nm excitation, however, no PL peaks associated with the divacancy were observed. In contrast, the pristine sample was studied under 915 nm excitation with a 1000 nm long-pass filter. Multiple spectra were recorded both from unprocessed bulk regions (reference) and from laser written spots at 1850 nJ, taken close to and far from markers, example spectra are shown in figure \ref{fig:lowtemp_NIR_Pristine}.

In the reference spectra from an unprocessed region, PL peaks characteristic of the divacancy are visible \cite{koehl_room_2011, magnusson_excitation_2018}: PL1 (1132 nm), PL2 (1130 nm), and PL4 (1078 nm). These centers were either present natively in the crystal or formed during the annealing step at 800°C. At the laser written spots, the same spectral features can be identified, however, they appear broadened compared to the reference. This broadening is attributed to lattice damage induced by the laser writing process, and is most pronounced around the alignment markers, where the crystal damage is most severe. The absence of any PL intensity enhancement and the spectral broadening of the divacancy-related peaks, suggests that the laser writing process did not generate additional divacancies, but instead degraded or partially destroyed the ones already present before laser processing.


We investigated the effect of epitaxial graphene on HPSI SiC on the creation of luminescent defects under femtosecond laser irradiation. We employ a commercial HPSI substrate and an industrial grade laser system to test broader applicability of the method. We observed a much lower intensity threshold to create PL defects on the sample with graphene compared to the pristine sample, suggesting graphene alters the local absorption and energy deposition conditions on the 4H-SiC surface during laser writing. In accordance with previous work, we do not observe efficient creation of $V_{Si}$ or divacancies centers, while spots with intense PL seem to be connected with strong surface structures (dimples) in a reference sample without graphene. Despite strong surface modification, we do not see signature of amorphous materials. To optimize the process, 4H-SiC substrate with a 4H-SiC epitaxial layer with a more favorable doping could be investigated as well as shorter laser wavelength and stronger focusing to enhance non-linear effects in comparison to sample heating.

\bibliography{aipsamp}

\begin{acknowledgments}
This work was funded by the Deutsche Forschungsgemeinschaft  (DFG, German Research Foundation)-- Project No. 429529648--TRR 306 QuCoLiMa ("Quantum Cooperativity of Light and Matter"). EN acknowledges support from the Quantum-Initiative Rhineland-Palatinate (QUIP). We thank Johannes Lehmeyer and Michael Krieger (FAU Erlangen, work-group H. Weber) for additional PL measurements of laser irradiated samples as well as the growth of the epitaxial graphene layer. 
FK acknowledges support by the Luxembourg National Research Fund (FNR) via the PEARL chair “AQuaTSiC” under grant agreement 17792569, as well as the European Research Council for the project “Q-Chip” under grant agreement 101171067.
FK and JH acknowledge further the European Union’s Horizon 2020 Research and Innovation Programme via the QuantERA project “SiCqurTech” under grant agreement 101017733, as well as the Horizon Europe Programme for the Flagship project “QIA” under grant agreement 101102140.
We thank Stefan Dix (RPTU, work-group A. Widera) and Thomas Utz (RPTU, work-group G. von Freymann) for additional experiments on laser writing and Konstantin Gröpl (RPTU, work-group V. Schünemann) for his help with the Raman measurements.

\end{acknowledgments}

\end{document}